\documentclass[fleqn,twoside]{article}
\usepackage{espcrc2,graphicx}

\title{Progress in multiloop calculations}

\author{A.G.~Grozin\address{Budker Institute of Nuclear Physics, Novosibirsk}}
       
\begin{document}

\begin{abstract}
I briefly summarize the talks on calculation of multiloop Feynman diagrams
presented at ACAT'2002 (Moscow University).
\vspace{1pc}
\end{abstract}

\maketitle

Here I briefly summarize the talks about multiloop
calculations~\cite{Ko:01,Gr:02,CCR:02,JKV:02,DS:02,FTRW:02,TF:02}
which were presented at the parallel session
``Simulations and Computations in Theoretical Physics and Phenomenology''
at the VIII International Workshop on Advanced Computing and Analysis Techniques
in Physics Research (ACAT'2002), June 24--28, 2002, Moscow.

In order to have some rough classification of these contributions,
we shall place them on the map (Table~\ref{Tab}) showing the number
of legs and loops.
The current state of the art is
\begin{itemize}
\item 0 legs --- 4 loops (with 1 mass);
\item 2 legs --- 3 loops (massless, HQET, on-shell);
\item 3, 4 legs --- 2 loops (various cases).
\end{itemize}

\begin{table}[h]
\caption{The map of multiloop calculations}
\label{Tab}
\begin{tabular}{|c|c|c|c|c|c}
\cline{1-5}
& 0 & 2 & 3 & 4 & legs \\
\cline{1-5}
1 & \textit{v1} & \textit{p1} & \textit{t1} & \textit{b1} & \\
2 & \textit{v2} & \textit{p2} & \textit{t2} & \textit{b2} & \\
3 & \textit{v3} & \textit{p3} &             &             & \\
4 & \textit{v4} &             &             &             & \\
\cline{1-5}
\multicolumn{1}{c}{loops} & \multicolumn{5}{c}{}
\end{tabular}
\end{table}

\section{Kotikov (\textit{p2})}

In his talk~\cite{Ko:01}, A.V.~Kotikov reviewed several methods of multiloop calculations.

\begin{figure}[h]
\begin{center}
\includegraphics{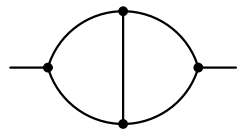}
\end{center}
\caption{Massless two-loop self-energy diagram.}
\label{F:K}
\end{figure}

First, he discussed application of the Gegenbauer polynomial technique
to the calculation of the massless two-loop diagram (Fig.~\ref{F:K})
with three non-integer powers of denominators.
The results can be expressed via ${}_3F_2$ hypergeometric functions
of the unit argument with indices tending to integers at $\varepsilon\to0$.
They can be expanded in $\varepsilon$ in terms of multiple $\zeta$-sums
(generalizing the Riemann function).
Reduction of such sums to a minimal set of independent ones
has been extensively investigated recently.
With these results, it is not difficult to expand these diagrams
up to rather high powers of $\varepsilon$.

Then he discussed the differential equations method.
All basis two-loop two-point diagrams with a single mass
(Fig.~\ref{F:K}, where some lines are massive)
have been calculated using this method.

\section{Grozin (\textit{p2}, \textit{p3})}

In his talk~\cite{Gr:02}, A.G.~Grozin reviewed the recent progress
in calculation of propagator diagrams in HQET and on-shell QCD.

It is well-known that there are 3 generic topologies
of massless 3-loop propagator diagrams.
They can be reduces, using integration by parts, to 6 basis integrals.
Four basis integrals are trivial.
One is the diagram of Fig.~\ref{F:K} with a non-integer power
of the middle line;
it is a particular case of a more general result
discussed in the previous talk.
The last and most difficult basis integral (the nonplanar one)
can be easily found at $\varepsilon=0$ by gluing;
there is no easy way to find further terms of its $\varepsilon$ expansion.

There are 10 generic topologies of 3-loop HQET propagator diagrams.
They can be reduced, using integration by parts, to 8 basis integrals.
Five basis integrals are trivial.
Two can be expressed via ${}_3F_2$ hypergeometric functions
of the unit argument;
they can be easily expanded to a rather high power in $\varepsilon$.
The last and most difficult basis integral was found
up to the finite term in $\varepsilon$,
using direct integration in the coordinate space.
More terms of its $\varepsilon$-expansion were recently obtained
using inversion.

There are 2 generic topologies of 2-loop on-shell QCD propagator diagrams
with a single non-zero mass.
They can be reduced, using integration by parts, to 3 basis integrals.
Two basis integrals are trivial, and the third one is expressed
via ${}_3F_2$ hypergeometric functions of the unit argument.
However, some of their indices tend to half-integers at $\varepsilon\to0$,
and the above algorithm of expansion in $\varepsilon$ is not applicable.
The case when there is another non-zero mass
was systematically studied recently.
There are 4 basis integrals, 2 of them trivial,
and 2 are expressed via ${}_3F_2$ hypergeometric functions
of the mass ratio squared.
Finite parts at $\varepsilon\to0$ are expressed via dilogarithms.

There are 11 generic topologies of 3-loop on-shell propagator diagrams
with a single non-zero mass.
They can be reduced, using integration by parts, to 18 basis integrals.
The basis integrals are mostly known from the calculation
of the electron magnetic moment in QED.
Some on-shell diagrams are related to HQET ones by inversion
of Euclidean integration momenta.

\section{Caffo (\textit{p2})}

\begin{figure}[h]
\vspace{9pt}
\centering
{\scalebox{.7}[.7]{\includegraphics*[80,20][370,120]{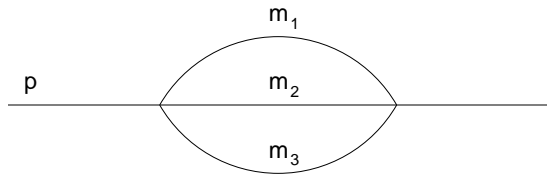}}}
\caption{The general massive 2-loop sunrise self-mass diagram.}
\label{F:C}
\end{figure}

In his talk~\cite{CCR:02}, M.~Caffo discussed numerical calculation
of the two-loop sunrise diagram with generic masses (Fig.~\ref{F:C})
for arbitrary $p^2$.
The finite parts of the basis integrals satisfy a system
of linear differential equations.
Their values at $p^2=0$, $\infty$, the threshold $(m_1+m_2+m_3)^2$,
and the pseudothresholds like $(m_1+m_2-m_3)^2$ are known.
The differential equations are integrated numerically,
starting from $p^2=0$, along a contour in the lower half-plane.

\section{Kalmykov (\textit{p2})}

In his talk~\cite{JKV:02}, M.Yu.~Kalmykov discussed pole masses
of gauge bosons in the Standard Model.
In order to relate them to the $\overline{\mathrm{MS}}$ masses
with the 2-loop accuracy,
one has to calculate their 1-loop self-energies and their derivatives,
and the 2-loop self-energies, on the mass shell.
After expansions in several small parameters,
most diagrams can be reduced to single-scale on-shell integrals,
which are expressed via basis integrals using integration by parts.
This is done by the FORM package ONSHELL2.
For diagrams with light fermion loops,
the expansion in $\sin\theta_W$ breaks down due to threshold singularities.
These diagrams are reduced to basis ones by Tarasov's recurrence relations;
in their sum, logarithms of $\sin\theta_W$ cancel.

\section{Davydychev (\textit{t1}, \textit{t2})}

In his talk~\cite{DS:02}, A.I.~Davydychev discussed the vertex diagrams
(Fig.~\ref{F:D}) on the mass shell ($P^2=M^2$, $p^2=m^2$).
The one-loop diagram can be exactly expressed via the one-loop
two-point diagram with masses $M$, $m$ and the momentum $q$
in $2-2\varepsilon$ dimensions.
The two-loop diagram has been calculated,
using Mellin-Barnes representation,
at $m=0$ and at non-zero $m\ll M$.

\newcommand{\triangleMm}
 {\setlength {\unitlength}{0.7mm}
 \begin{picture}(36,20)(0,28)
\thicklines
 \put (18,48) {\line(0,1){6}}
 \put (18,48) {\line(-1,-3){12}}
\thinlines
 \put (18,48) {\line(1,-3){12}}
 \multiput(8.5,18)(2,0){10}{\line(1,0){1}}
 \put (18,48) {\circle*{1}}   
 \put (8,18)  {\circle*{1}}
 \put (28,18) {\circle*{1}}
 \put (6,30)  {\makebox(0,0)[bl]{\large $M$}}
 \put (26,30) {\makebox(0,0)[bl]{\large $m$}}
 \put (17,13)  {\makebox(0,0)[bl]{\large $0$}}
 \put (0,12) {\makebox(0,0)[bl]{\large $P$}}
 \put (32,11) {\makebox(0,0)[bl]{\large $p$}}
 \put (14,50) {\makebox(0,0)[bl]{\large $q$}}   
 \end{picture}}
 
\newcommand{\vertexMm}
 {\setlength {\unitlength}{0.7mm}
 \begin{picture}(36,20)(0,28)
\thicklines
 \put (18,48) {\line(0,1){6}}
 \put (18,48) {\line(-1,-3){12}}
\thinlines
 \put (18,48) {\line(1,-3){12}}
 \multiput(13.5,33)(2,0){5}{\line(1,0){1}}
 \multiput(8.5,18)(2,0){10}{\line(1,0){1}}
 \put (18,48) {\circle*{1}}   
 \put (13,33)  {\circle*{1}}
 \put (23,33) {\circle*{1}}
 \put (8,18)  {\circle*{1}}
 \put (28,18) {\circle*{1}}
 \put (4,25)  {\makebox(0,0)[bl]{\large $M$}}
 \put (27,25) {\makebox(0,0)[bl]{\large $m$}}
 \put (9,40)  {\makebox(0,0)[bl]{\large $M$}}
 \put (22,40) {\makebox(0,0)[bl]{\large $m$}}
 \put (17,13)  {\makebox(0,0)[bl]{\large $0$}}
 \put (17,28) {\makebox(0,0)[bl]{\large $0$}}
 \put (0,12) {\makebox(0,0)[bl]{\large $P$}}
 \put (32,11) {\makebox(0,0)[bl]{\large $p$}}
 \put (14,50) {\makebox(0,0)[bl]{\large $q$}}   
 \end{picture}}

\begin{figure}[h]
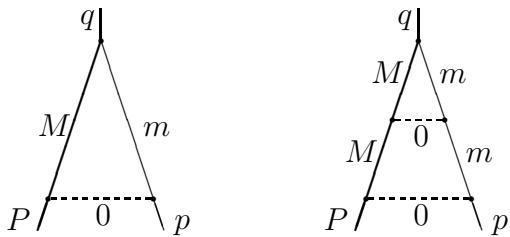

\[
\begin{array}{ccc}
\triangleMm & $\hspace{1cm}$ & \vertexMm
\end{array}
\]
\caption{One- and two-loop vertex diagrams}
\label{F:D}
\end{figure}

\section{Fleischer (\textit{t1}, \textit{b1})}

In his talk~\cite{FTRW:02}, J.~Fleischer discussed some steps towards
the two-loop Bhabha scattering calculation.
The matrix element is decomposed into Dirac structures
times scalar amplitudes.
One-loop amplitudes,
as well as two-loop amplitudes with the electron mass counterterm
(they are reduce to one-loop integrals),
have been calculated.
Tensor integrals are reduced to scalar ones with shifted dimension.
Scalar integrals are reduced to basis ones.
One-loop on-shell trianglular diagrams are reduced to two-point ones
(as in the previous talk).

\section{Tentyukov}

In his talk~\cite{TF:02}, M.~Tentyukov presented the DIagram ANAlizer (DIANA).
It does not calculate diagrams;
it does other things necessary for large-scale automatic diagram calculations.
Diagrams for a given process can be generated by QGRAF.
But calculation of diagrams of each topology must follow a separate path.
DIANA sorts generated diagrams into topologies.
Packages for calculation of Feynman integrals (such as Mincer)
require a specific routing of external and loop momenta for each topology.
QGRAF provides some momentum routing for each diagram,
but not the required one.
DIANA corrects this deficiency.
Last but not least, it can produce nice graphical representations
of diagrams, ready for inclusion into a paper.

\end{document}